\documentstyle[12pt,fleqn]{article}

\def\f{\phi}
\def\fb{\bar {\phi}}
\def\ff{\phi_{\rm F}}

\catcode`@=11
\def\@versim#1#2{\lower.7\p@\vbox{\baselineskip\z@skip\lineskip-.5\p@
    \ialign{$\m@th#1\hfil##\hfil$\crcr#2\crcr\sim\crcr}}}
\def\ga{\mathrel{\mathpalette\@versim> }}
\def\la{\mathrel{\mathpalette\@versim<}}
\catcode`@=12

\vspace{0.5in}

\title{Thermal Fluctuations and Validity of the 1-loop Effective Potential}

\author{Marcelo Gleiser\thanks{gleiser@peterpan.dartmouth.edu} $\:$  and
Rudnei O. Ramos\thanks{rudnei@northstar.dartmouth.edu} \\
{\it Department of Physics and Astronomy, Dartmouth College}\\
{\it Hanover, NH 03755}}

\date{DART-HEP-92/08 \\ November 1992}

\begin{document}

\maketitle

\begin{abstract}

We examine the validity of the 1-loop approximation to the effective potential
at finite temperatures and
present a simple
test for its reliability. As an application we study the
standard electroweak potential, showing that for a Higgs mass above
$70$ GeV, and fairly independently of the top mass (with $m_t\ge 90$ GeV),
the 1-loop approximation
is no longer valid for temperatures in the neighborhood of
the critical temperature.


\end{abstract}

\section{Introduction}

The effective potential at
finite temperatures is an important tool in the study of phase
transitions in scalar and gauge field theories [\ref{r:1}].
It is equivalent to the homogeneous coarse-grained free-energy
density functional of
statistical
physics, with its minima giving the stable and, when applicable, metastable
states of the system.
For interacting field theories the effective potential is evaluated
perturbatively, with an expansion in loops being equivalent to an expansion
in powers of $\hbar$ [\ref{r:CW}]. The 1-loop approximation is then
equivalent to incorporating the first quantum corrections to the classical
potential. We start by briefly reviewing the calculation of the 1-loop
potential for a self-interacting scalar field theory.
The classical action in the presence of an external source $J(x)$ is
\begin{equation}
S[\phi,J]=\int d^4x \left [{1\over 2}\partial_{\mu}\phi\partial^{\mu}\phi
-V(\phi)+\hbar J(x)\phi(x)\right ] \:.
\label{e:1}
\end{equation}
The effective action $\Gamma[\phi_c]$ is defined in terms of the
connected generating functional $W[J]$ as
\begin{equation}
\Gamma[\phi_c]=W[J]-\int d^4x J(x)\phi_c(x) \:~~,
\label{e:2}
\end{equation}
where the classical field $\phi_c(\vec{x},t)$ is defined by
$\phi_c(\vec{x},t)\equiv
\delta W[J]/
\delta J(x)$, and
\begin{equation}
W[J]=-i\hbar {\rm ln}\int D \phi ~ {\rm exp}\left [{i\over {\hbar}}S[\phi,J]
\right ] \:.
\label{e:3}
\end{equation}

In order to evaluate $\Gamma[\phi_c]$ perturbatively, one writes
the field as
$\phi(\vec{x},t) \rightarrow \phi_{0}(\vec{x},t) + \eta(\vec{x},t)$, where
$\phi_{0}(\vec{x},t)$ is a field configuration which extremizes
the classical action
$S[\phi,J]$, $\frac{\delta S[\phi,J]}{\delta \phi} |_{\phi =
\phi_{0}} = 0$,
and
$\eta ({\vec{x},t})$ is a small perturbation about that extremum configuration.
The action $S[\phi,J]$ can then be expanded about $\phi_0(\vec{x},t)$
and, up to quadratic order in $\eta(\vec{x},t)$, we can use a saddle-point
approximation to the path integral to obtain
for the connected generating functional,
\begin{equation}
W[J]=S[\phi_0]+\hbar\int d^4x \phi_0(x)J(x)+{{i\hbar}\over 2}
{\rm Tr ln}\left [\partial_{\mu}\partial^{\mu}+V^{\prime \prime}
(\phi_0)\right ] \:.
\label{eq:4}
\end{equation}
In order to obtain the 1-loop expression for $\Gamma[\phi_c]$, we first note
that writing $\phi_0=\phi_c-\eta$ we get to first order in $\hbar$,
$S[\phi_0]=S[\phi_c]-\hbar\int d^4x \eta(x)J(x)+O(\hbar^2)$. Using this
result and
Eq. (4) into Eq. (2) we find, as $J\rightarrow 0$,
\begin{equation}
\Gamma[\phi_c]=S[\phi_c]+{{i\hbar}\over 2}{\rm Tr ln}
\left [\partial_{\mu}\partial^{\mu}+V^{\prime \prime}
(\phi_c)\right ] \:.
\label{eq:5}
\end{equation}
The effective action can also be computed as a derivative expansion about
$\phi_c(\vec{x},t)$,
\begin{equation}
\Gamma[\phi_c]=\int d^4x\left [-V_{\rm eff}(\phi_c(x))+{1\over 2}
\left (\partial_{\mu}
\phi_c\right )^2Z(\phi_c(x))+\dots \right ] \:.
\label{e:6}
\end{equation}
The function $V_{\rm eff}(\phi_c)$ is the
effective potential. For a constant field
configuration $\phi_c(\vec{x},t)=\phi_c$ we obtain
\begin{equation}
\Gamma[\phi_c]=-\Omega V_{\rm eff}(\phi_c) \:,
\label{eq:7}
\end{equation}
where $\Omega$ is the total volume of space-time. Comparing Eqs. (5) and (7) we
obtain for the 1-loop effective potential,
\begin{equation}
V_{\rm eff}(\phi_c)=V(\phi_c)-{{i\hbar}\over 2}\Omega^{-1}{\rm Tr ln}
\left [\partial_{\mu}\partial^{\mu}+V^{\prime \prime}
(\phi_c)\right ] \:.
\label{eq:8}
\end{equation}

When working at non-vanishing temperature, the same functional techniques can
be used. In this case one is interested in evaluating the generating
functional (the partition function)
$Z_{\beta}[J]$ which is given by the path integral [\ref{r:KAPUS}]
\begin{equation}
Z_{\beta}[J]=N\int D\phi {\rm exp}\left [-\int_0^{\beta}d\tau \int d^3x\left
({\cal L}_E-J\phi\right )\right ] \:,
\label{eq:9}
\end{equation}
where the integration is restricted to paths periodic in $\tau$ with
$\phi(0,\vec{x})=\phi(\beta,\vec{x})$, ${\cal L}_E$ is the Euclidean
Lagrangian, and $N$ is a normalization constant. Again one expands
about an extremum of the Euclidean action and calculates the partition
function by a saddle-point evaluation of the path integral. The result for
the 1-loop approximation to the effective potential is
\begin{equation}
V_{\rm eff}(\phi_c,T)=V_{\rm eff}(\phi_c)+{{\hbar}\over {2\pi^2\beta^4}}
\int_0^{\infty}dx~x^2{\rm ln}\left \{1-{\rm exp}\left [-\sqrt{x^2+\beta^2
V^{\prime \prime}(\phi_c)}\right ] \right \} \:~.
\label{eq:10}
\end{equation}

 From the above discussion it is clear that the 1-loop approximation to
the effective action, Eq. (5), works best when the classical field does not
differ much from the configuration that extremizes the classical action,
$\phi_c=\phi_0 +\eta\sim \phi_0$, since in this case the saddle-point
evaluation to the path integral is adequate.  Also, $\f_c(\vec{x},t)$
must be nearly
constant so that the effective potential can be obtained from Eq.
(7). As $J\rightarrow 0,~\f_c(\vec{x},t)$ is identified with
$\langle\f\rangle$, the vacuum expectation value.
How large can be the fluctuation
$\eta(\vec{x},t)$ without spoiling the validity of the approximation? Clearly,
for models that exhibit a second order transition, the approximation
worsens as one approaches the critical temperature from above or below, with
infrared corrections becoming progressively more important.
One way of dealing with this problem is to obtain an
improved effective potential where some of the infrared divergences
are taken into account, for example by the summation of
daisy (or super daisy) diagrams [\ref{r:DJ}]. This method has recently been
extensively discussed in connection with the standard electroweak potential,
in an attempt to include infrared effects from higher gauge loops
[\ref{r:DAISY}]. (For vector boson masses $m_{\rm V}\sim g\f$, the expansion
parameter $g^2T/m_{\rm V}\sim gT/\f$ is large for small values of $\f$.)
Another approach is to use
$\varepsilon$-expansion techniques in order to compute corrections to the
critical exponents that control the singular behavior of physical quantities
near the critical point, as is familiar from the theory of critical
phenomena [\ref{r:EXP}]. Here, we will not be concerned with improving the
1-loop approximation, but in quantifying its reliability. Our results should be
of relevance in particular in the study of weakly first-order transitions,
where large fluctuations about equilibrium may be present, invalidating the
1-loop approximation for certain values of the temperature or other relevant
physical parameters of $V_{\rm eff}(\phi,T)$.

In this paper we propose a simple method to estimate the validity of the
1-loop approximation to the effective potential. We will argue that the
statistically dominant thermal fluctuations around the minimum of
$V_{\rm eff}(\phi,T)$ are spherically symmetric and
have roughly a correlation volume, where the correlation length is given
by the inverse temperature dependent mass, $\xi(T)=m^{-1}(T)$. Assuming that
these fluctuations are Boltzmann suppressed, we
compute the average value for their amplitude. Following Gleiser, Kolb, and
Watkins (GKW) [\ref{r:GKW}] we will refer
to these fluctuations as sub-critical bubbles. Note, however, that the
sub-critical bubbles of GKW had fixed amplitude, while here we will
average over all possible fluctuations. Contrary to GKW we are not
interested in the dynamics of the transition, but on the validity of the
1-loop potential. (This also explains our emphasis on the effective potential
as opposed to the effective action.)

For small enough amplitudes
the 1-loop approximation clearly is satisfactory. Otherwise, infrared
corrections are important, and the 1-loop approximation is unreliable.
In the next Section we obtain the free energy of correlation volume
thermal fluctuations. In Section III we discuss the
validity of the 1-loop finite temperature effective
potential in the presence of sub-critical thermal fluctuations. In
Section IV we apply our results to the electroweak potential.
Conclusions are presented in Section V.

\section{Free Energy of Thermal Fluctuations}

The idea that the statistically dominant fluctuations around equilibrium
can be modelled by sub-critical bubbles of roughly a correlation volume has
been discussed by GKW and other recent works [\ref{r:BUBII},\ref
{r:DINE}].
Although
the original proposal of GKW, that sub-critical bubbles of the broken-symmetric
phase
can play an important r\^ole in the dynamics of weakly first-order
transitions is still under debate [\ref{r:DINE}], their presence in any
hot fluctuating system is undisputed. In statistical physics, the
coarse-grained free energy functional is built under the assumption that the
relevant coarse-graining scale is the correlation length, $\xi(T)$;
if the
coarse-graining scale were to be larger than the correlation length, phase
separation would occur within a ``grain'' and the free energy would become a
convex function of the order parameter [\ref{r:LANGER}].
Thus, the coarse-graining
guarantees that we can identify different phases in our system, so
long as the fluctuations about equilibrium are consistent with the
coarse-graining scale.
The same reasoning applies to the 1-loop approximation to the effective
potential. The effective coarse-graining scale is given by the inverse mass of
fluctuations about the equilibrium state, the correlation length $\xi(T)=
m^{-1}(T)$. If we want to use
the effective potential to describe a first-order phase transition it
better be a
concave function of the scalar field at, say,  the critical temperature.
Hence, the value of $\phi$ inside the correlation volume fluctuations
should not differ much from its equilibrium value.

We now briefly estimate the free energy of sub-critical fluctuations. We refer
the reader to GKW for details. Although our approach is quite general, it is
adequate to perform the calculation for a particular potential
which in principle includes interactions of $\f$ with itself and other
fields.
We write the 1-loop effective potential as

\begin{equation}
V_{\rm eff}(\phi,T) = \frac{m^{2}(T)}{2} \phi^{2} - \gamma(T) \phi^{3} +
\frac{\lambda(T)}{4} \phi^{4}  \: ,
\label{e:pot}
\end{equation}

\noindent
where $\gamma(T)$ and $\lambda(T)$ are positive functions of $T$ and
$m^{2}(T)$ can be negative below a certain temperature $T_{2} < T_{c}$.
$V_{\rm eff}(\phi,T)$ has minima at $\phi=0$ (for $ m^{2}(T) > 0$) and
at $\phi_{+} = \frac{1}{2 \lambda(T)} \left [ 3 \gamma(T) +
\sqrt{ 9 \gamma^{2}(T)
- 4 m^{2}(T) \lambda(T)} ~\right ]$, for temperatures $T < T_{1}$, with $T_{1}$
given by the solution of $\gamma^{2}(T_{1}) = \frac{4}{9} m^{2}(T_{1})
\lambda(T_{1})$. At $T= T_{c}$, $V_{\rm eff}(\phi=0,T_{c}) = V_{\rm eff}(\phi=
\phi_{+}, T_{c})$. Below $T_{c}$ the minimum at $\phi= \phi_{+}$ becomes the
global minimum (the true vacuum) and the minimum at $\phi=0$ becomes
metastable (the false vacuum).
Note that $\phi$ could be a real scalar field or the amplitude of the Higgs
field. In the latter case, $V_{\rm eff}$ is an even function of $\phi$.

Consider cooling the system described by the above potential from $T\gg T_c$
down to $T\ga T_c$. The equilibrium state of the system is at $\phi=0$.
The probability of a thermal fluctuation $\ff(\vec{x})$ about $\phi=0$ is

\begin{equation}
P[\ff] \sim {\rm exp}\left [{-\frac{F(\ff,T)}{T}}\right ] \: ,
\label{e:prob}
\end{equation}

\noindent
where $F(\ff ,T)$ denotes the excess
free energy of the thermal fluctuation.
Following
Ref. [\ref{r:GKW}] we write $F(\ff,T)$ as

\begin{equation}
F(\ff,T) = \int dV \left[ \frac{1}{2} (\vec{\nabla} \ff)^{2} +
V_{\rm eff}(\ff,T) \right] \: .
\label{e:13}
\end{equation}
Note that the free energy is equivalent to the (Euclidean) effective action,
as defined
in Eq. (\ref{e:6}), to first order in the
derivatives (with $Z(\f_c(\vec{x}))=1$), for a static field configuration.
One could improve on this approximation by including higher order terms in
$F(\ff,T)$, although we refrain from doing so here [\ref{r:EFFEC}]. How can
we estimate the free energy of these fluctuations? Since they are not
extrema of the classical action (like, {\it e.g.}, critical bubbles) we must
choose an explicit profile for the
typical fluctuations. In order to minimize the free energy we choose them
to be spherically symmetric. They can then be described by two parameters,
their
radius and the value of the field $\phi$ in their interior, $\f_A$. (We refer
to $\f_A$ as the amplitude of the fluctuation.)
Following the
discussion above we take the radius to be the temperature dependent
correlation radius and write
\begin{equation}
\ff (r,T)  = \f_A \exp \left( - \frac{r^{2}}{\xi(T)^{2}} \right) \: .
\label{e:bubble}
\end{equation}
\noindent
Other choices for $\ff (r,T)$ give larger free energy. The free energy
for the correlation volume fluctuations with amplitude $\f_A$
becomes
\begin{equation}
F(\f_A,T) = \alpha(\f_A) \xi(T) + \beta(\f_A,T) \xi(T)^{3} \: ,
\label{e:15}
\end{equation}
\noindent
where $\alpha(\f_A)$ and $\beta(\f_A,T)$ are given, respectively, by
\[
\alpha(\f_A) = \frac{3 \sqrt{2}}{8} \pi^{\frac{3}{2}} \f_A^{2} \:,
\]
\begin{equation}
\label{e:alpha}
\end{equation}
\[
\beta(\f_A,T) = \frac{\sqrt{2}}{8} \pi^{\frac{3}{2}} m^{2}(T)
\phi_A^{2} - \frac{\sqrt{3}}{9} \pi^{\frac{3}{2}} \gamma(T)
\phi_A^{3} + \frac{\pi^{\frac{3}{2}}}{32} \lambda(T) \phi_A^{4} \: .
\]

\section{Validity of the 1-loop Aproximation}

As we discussed in the Introduction, the 1-loop approximation to the effective
action is obtained by expanding the classical action about an extremum
configuration, $\f_0(\vec{x},t)$,
and keeping terms up to second order in the perturbation
$\eta(\vec{x},t)$, with
the classical field $\f_c(\vec{x},t)=\f_0(\vec{x},t)+\eta(\vec{x},t)$.
When $J(x)
\rightarrow 0$, $\f_c(\vec{x},t)$ becomes a constant, the vacuum expectation
value $\langle \f\rangle$, which, by Eq. (7),
is a solution of $dV_{\rm eff}(\f_c)/d\f_c |_{
\langle \f\rangle} = 0$. Thus, the 1-loop approximation to the effective
potential relies on having
fluctuations about $\f_c=\langle\f\rangle$ which are small enough that the
inhomogeneous terms in the effective action (Eq. (6)) can
be neglected. For the models
described by the potential of Eq. (\ref{e:pot}), for $T\ga T_c$,  we
are interested in the amplitude of fluctuations about $\f_c=0$.
For $T<T_1$, $V_{\rm eff}
(\f,T)$ has an inflexion point closest to $\f_c=0$ at
\begin{equation}
\f_{\rm inf}(T)={{\gamma(T)}\over {\lambda(T)}}-\sqrt{{{\gamma^2(T)}\over
{\lambda^2(T)}}-{{m^2(T)}\over {3\lambda(T)}} } \:.
\label{e:inf}
\end{equation}
Clearly, the rms amplitude of fluctuations, which we write as $\fb(T)$,
must be
smaller than $\f_{\rm inf}(T)$
in order for the 1-loop approximation to be accurate.
Thus we can write as a criterion for the
validity of the 1-loop approximation,
\begin{equation}
\fb(T) \le \f_{\rm inf}(T) \:.
\label{e:crit}
\end{equation}
This is a general criterion which can be adapted to different models,
including second-order transitions in the neighborhood of the critical point.

What remains is to calculate $\fb(T)$ [\ref{r:MG}].
Since $\fb(T)$ is the rms amplitude
of the fluctuations, its definition is simply,
\begin{equation}
\fb(T)=\sqrt{\langle\f^2\rangle_T - \langle\f\rangle^2_T } \:,
\label{e:rms}
\end{equation}
where the thermal average $\langle\dots\rangle_T$ is defined in terms of the
probability distribution of Eq. \ref{e:prob} as
\begin{equation}
\langle\dots\rangle_T={{\int_{-\infty}^{+\infty}D\f\dots P[\f]}\over
{\int_{-\infty}^{+\infty}D\f~P[\f]}} \:.
\label{e:average}
\end{equation}
Note that with our ansatz of Eq. (\ref{e:bubble}) for the thermal
fluctuations,
the path integrals above become simple integrals over $\f_A$.

\section{Application: The Electroweak 1-loop Potential}

As an application we study the 1-loop approximation to the electroweak
potential given by [\ref{r:AH}]
\begin{equation}
V_{\rm eff}(\phi ,T) = D (T^{2} - T_{2}^{2}) \phi^{2} - ET \phi^{3} +
\frac{\lambda_{T}}{4} \phi^{4} \: ,
\label{e:vew}
\end{equation}

\noindent
where $D$ and $E$ are constants given in terms of the $W$ and $Z$ boson masses
and of the top quark mass as $D = \frac{1}{24} \left[ 6 \left(
\frac{m_{W}}{\sigma} \right)^{2} + 3 \left(\frac{m_{Z}}{\sigma} \right)^{2} +
6 \left(\frac{m_{t}}{\sigma} \right)^{2} \right]$  and $ E = \frac{1}{12 \pi}
\left[ 6 \left(\frac{m_{W}}{\sigma} \right)^{3} + 3 \left(\frac{m_{Z}}{\sigma}
\right)^{3} \right] \simeq 10^{-2}$, where $\sigma \simeq 246$ GeV
is the vacuum expectation value
of the Higgs field. We use $m_W=80.6$ GeV and $m_Z=91.2$ GeV. $T_{2}$
is the spinodal instability temperature, given by
\begin{equation}
T_{2} = \sqrt{ \frac{ m_{H}^{2} - 8 B \sigma^{2}}{ 4 D}} \: ,
\label{e:T2}
\end{equation}

\noindent
where $m_{H}^{2} = \frac{2 \lambda + 12 B}{ \sigma^{2}}$ is the physical
Higgs mass and $B= \frac{1}{64 \pi^{2} \sigma^{4}} \left( 6 m_{W}^{4} +
3 m_{Z}^{4} - 12 m_{t}^{4} \right)$. The temperature dependent Higgs
self-coupling $\lambda_{T}$ is given by
\begin{equation}
\lambda_{T} = \lambda - \frac{1}{16 \pi^{2}} \left[ \sum_{b} g_{b}
\left( \frac{m_{b}}{\sigma} \right)^{4} \ln \left( \frac{ m_{b}^{2}}{ c_{b}
T^{2}} \right) - \sum_{f} g_{f} \left(\frac{m_{f}}{\sigma} \right)^{4}
\ln \left( \frac{ m_{f}^{2}}{ c_{f}T^{2}} \right) \right] \:,
\label{e:lT}
\end{equation}
\noindent
where the sums are performed over bosons and fermions, with degrees of
freedom $g_{b}$ and $g_{f}$, respectively. In Eq. (\ref{e:lT}), $\ln c_{b} =
5.41$ and $\ln c_{f} = 2.64$.

The electroweak potential is equivalent to the potential of Eq. (\ref{e:pot}),
with the identifications $m^2(T)=2D(T^2-T_2^2),~\gamma(T)=ET,$ and $\lambda(T)=
\lambda_T$. At $T_c$ the minima at $\f=0$ and $\f_+$ are degenerate, with
\begin{equation}
T_{c}^{2} = {{T_{2}^{2}}\over { 1 - \frac{ E^{2}}{\lambda_{T} D}}} \:.
\label{e:Tc}
\end{equation}
\noindent
For $T< T_{1}$, the nearest inflexion
point to the minimum $\phi=0$ is located at
\begin{equation}
\phi_{\rm inf}(T) = \frac{ET}{\lambda_{T}} - \sqrt{ \frac{E^{2}T^{2}}{
\lambda_{T}^{2}} - \frac{2 D (T^{2} - T_{2}^{2} )}{3 \lambda_{T}} } \:.
\label{e:19}
\end{equation}
\noindent
It is now simple to obtain the expression for $\fb(T)$. From Eqs.
(\ref{e:prob}),
(\ref{e:alpha}), and (\ref{e:average}) we obtain (in the electroweak model
the potential is left-right symmetric, and $\langle\f_A\rangle_T=0$)
\begin{equation}
\left [\fb(T)\right ]^2 = {{ \int_{-\infty}^{\infty} d \phi_A \:\: \phi_A^2
e^{-\frac{1}{T} \left [ \alpha(\phi_A)\xi + \beta(\f_A,T)\xi^3 \right ]} }
\over {
\int_{-\infty}^{\infty} d \phi_A e^{-\frac{1}{T} \left
[ \alpha(\phi_A)\xi + \beta(\phi_A,T)\xi^{3} \right ]}} } \:.
\label{e:fb}
\end{equation}

\noindent
where
$\xi^{-1}(T) = \sqrt{ 2 D (T^{2} - T_{2}^{2})}$. Due to the non-linear terms
the integrals above cannot be calculated exactly. However, for the case
at hand,
the free energy of the fluctuations is dominated by their surface term. We
can safely set $\beta(\f_A,T)=0$ in the integrals above to obtain an
approximate analytic expression for $\fb(T)$,
\begin{equation}
\fb(T)\simeq \left[ \frac{ 4 D^{\frac{1}{2}} T (T^{2} -
T_{2}^{2})^{\frac{1}{2}} }{3 \pi^{\frac{3}{2}}} \right]^{\frac{1}{2}} \:.
\label{e:ftheo}
\end{equation}
\noindent
This result can be written as $\fb^2(T)\simeq m(T)T/6$ [\ref{r:MG}].
In Fig. 1 we compare, at the critical temperature,
the analytical result above for $\fb(T)$ with
the numerical result obtained by keeping the volume contribution to the
free energy. To each pair of curves corresponds a top mass. Within each pair,
the top curve is the numerical result, while the bottom curve is the
approximation of Eq. (\ref{e:ftheo}).
It is
clear that the approximation is very good, working to within $10\%$ at its
worse for all values of the Higgs mass
we investigated, being also only weakly dependent on the top mass.

The condition for the validity of the 1-loop approximation, Eq. (\ref{e:crit}),
reads,
\begin{equation}
\left[ \frac{ 4 D^{\frac{1}{2}} T (T^{2} -
T_{2}^{2})^{\frac{1}{2}} }{3 \pi^{\frac{3}{2}}} \right]^{\frac{1}{2}} \leq
\frac{ET}{\lambda_{T}} - \sqrt{ \frac{E^{2}T^{2}}{
\lambda_{T}^{2}} - \frac{2 D (T^{2} - T_{2}^{2} )}{3 \lambda_{T}} } \:.
\label{e:critew}
\end{equation}
This condition can be studied in two ways, assuming the results are fairly
independent of the top mass. (Or for a fixed top mass.)
We can either fix the Higgs mass
and look for the temperature that violates the inequality, or fix the
temperature and look for the Higgs mass that violates the inequality. We choose
the latter approach and look for the Higgs mass that violates the inequality
at the critical temperature. If $V_{\rm eff}(\f,T)$ is not a good approximation
at $T_c$ it should not be trusted for any temperatures $T_c\leq T\leq T_2$.
Solving
for $\lambda_T$ we obtain, using Eq. (\ref{e:Tc}),
\begin{equation}
\lambda_{T} \leq \pi\left [ E\left (1-{{\sqrt{3}}\over 2}\right )
\right ]^{2/3} \:.
\label{e:result}
\end{equation}
\noindent
In order to express this result in terms of the Higgs mass, note that at
$T_c$ we can write [\ref{r:GK}], (effectively approximating $\lambda_T$
to its tree-level value, $\lambda=m_H^2/2\sigma^2$)
\begin{equation}
\lambda_{T} \simeq 0.08 \left( \frac{m_{H}}{100GeV} \right)^{2} \:.
\label{e:lT2}
\end{equation}
\noindent
Substituting Eq. (\ref{e:lT2}) into Eq. (\ref{e:result}), we find
that for a Higgs mass $m_H\ga 70$ GeV the 1-loop approximation is no
longer valid. (Without the approximation
$\lambda_T=\lambda$ we find numerically $m_H\ga 77$ GeV.)
In Fig. 2 we compare this approximate analytical result with
the numerical result obtained by keeping all terms in the free energy. The
results turn out to be quite independent of the top mass, and in very good
agreement with each other.

\section{Conclusions}

We have examined the validity of the 1-loop approximation to the effective
potential at finite temperature. By modelling the statistically dominant
thermal fluctuations about equilibrium by correlation volume sub-critical
bubbles of arbitrary amplitude, we argued that the 1-loop potential is
valid so long as the rms amplitude of the fluctuations is smaller than the
closest inflexion point.

We applied our results to the electroweak model,
showing
analytically and numerically that for temperatures at and below the
critical temperature the 1-loop approximation breaks
down for Higgs masses $m_H\ga 70$ GeV. The results
depend only very weakly on the top mass. For smaller
Higgs masses, it is possible to trust the 1-loop approximation for
temperatures below $T_c$. In a more detailed study, it would be interesting
to obtain the maximum value of $m_H$ for which the potential is still valid
at the nucleation temperature for critical bubbles. Given that the
experimental lower
bound on the Higgs mass is $m_H\ga 60$ GeV, we suspect that the 1-loop
approximation will be ruled out for all values of $m_H$.

In closing, we mention that the perturbation expansion parameter for
scalar loops, $\lambda_TT/m(T)$, when evaluated at $T_c$, becomes bigger
than unity for $m_H\ga 85$ GeV [\ref{r:DAISY},\ref{r:GK}].
It is reassuring to note that our results are in qualitative
agreement with these power counting perturbative arguments, even though they
are non-perturbative in nature.

\vskip 1.cm

\centerline{\bf Acknowledgements}

\vskip 0.25cm
We would like to thank A. Linde for important comments and discussions.
(MG) is partially supported by a National Science Foundation grant No.
PHY-9204726. (ROR) is supported by a grant from Conselho Nacional de
Desenvolvimento Cient\'{\i}fico e Tecnol\'ogico - CNPq (Brazil).

\newpage

\vskip 2.cm
\centerline{\bf Figure Captions}

\noindent
Figure 1: Comparison between analytical
and numerical results for the rms
fluctuation amplitude $\fb(T)$ at the critical temperature as a
function of the Higgs mass. Each pair of curves is for a value of the
top mass. Within each pair, the top curve is the numerical result and the
bottom curve the approximation of Eq. (\ref{e:fb}).

\noindent
Figure 2: Criterion for the validity of the 1-loop approximation for the
electroweak potential at the critical temperature
as a function of the Higgs mass, for several values of the top mass. The
ascending curves are the numerical results for the
rms amplitude $\fb(T)$, while the descending curves give the location of
the inflexion point.
Values of $m_H$ to the right of the dots violate the inequality in
Eq. (18).

\end{document}